\documentclass[setspace,amsmath,amssymb]{article}
\usepackage{setspace}
\usepackage{graphicx}
\usepackage{amsmath}
\usepackage{amssymb}
 
\newcommand{\ra}{\rangle}
\newcommand{\la}{\langle}
\newcommand{\rra}{\rangle\!\rangle}
\newcommand{\lla}{\langle\!\langle}
\begin{document}
\begin{titlepage}

\begin{flushright}
\today
\end{flushright}

\vspace{1in}

\begin{center}

{\bf On-shell equation of the Lorentzian classicalized holographic tensor network}

\vspace{1in}

\normalsize

{Eiji Konishi\footnote{E-mail address: konishi.eiji.27c@kyoto-u.jp}}

\normalsize
\vspace{.5in}

{\it Graduate School of Human and Environmental Studies,\\
 Kyoto University, Kyoto 606-8501, Japan}
\end{center}

\vspace{1in}

\baselineskip=24pt
\begin{abstract}
In the Lorentzian classicalized holographic tensor network (cHTN), we derive its relativistic on-shell equation from its Lorentzian action in the presence of a relativistic massive particle in the bulk spacetime: $-\sigma \hbar \theta=Mc^2$.
Here, $\sigma$ is the von Neumann entropy of the cHTN per site in nats, $\theta$ is the real-proper-time expansion of the cHTN defined along the world line of the particle, and $M$ is the non-zero mass of the particle.
We explain the physical properties, interpretation, and consequences of this equation.
Specifically, from this equation we derive the properties of the on-shell proper acceleration of another massive particle in the bulk spacetime as those of the gravitational acceleration induced by the original massive particle.
\end{abstract}

\vspace{.1in}


\vspace{.6in}

\end{titlepage}

The known example of the holographic principle \cite{Hol1,Hol2,Hol3} is the $d+2$-dimensional anti-de Sitter spacetime/$d+1$-dimensional conformal field theory (AdS$_{d+2}$/CFT$_{d+1}$) correspondence \cite{AdSCFT1,AdSCFT2,AdSCFT3,AdSCFT4,Nastase}.
In the AdS$_3$/CFT$_2$ correspondence, the holographic tensor network (HTN) was initially proposed by Swingle in Ref. \cite{Swingle1} as the multi-scale entanglement renormalization ansatz (MERA) \cite{Vidal1,Vidal2} of the ground state of the strongly coupled boundary CFT$_2$ and corroborates the Ryu--Takayanagi formula for the holographic entanglement entropy defined for this boundary quantum state \cite{RT1,RT2,HRT,RT3,Review0}.

Based on this HTN proposal \cite{Swingle1,Matsueda,Review1,Swingle2,Review3}, we classicalize the HTN (i.e., the MERA) of the boundary CFT$_2$ from a quantum pure state to its completely mixed quantum state by restricting the set of observables for the qubit Hilbert space, ${\mathfrak H}$, of the boundary CFT$_2$ to an Abelian subset, ${\cal A}$, as in Refs. \cite{EPL1,EPL2,JHAP1}.
This restriction of the set of observables is done by introducing a superselection rule operator, that is, the Pauli third matrix, which is diagonal in the qubit eigenbasis.
Here, the elements of ${\cal A}$ are defined by the commutativity with the superselection rule operator, and thus the quantum interferences of the HTN in the qubit eigenbasis are completely lost.

Note that here we use the term {\it classicalization} in the sense of the {\it quantum-to-classical transition} in quantum decoherence theory \cite{ZurekPT}.
The state obtained from a coherent quantum pure state by classicalization is a mixed {\it quantum} state and belongs not to classical mechanics but to quantum mechanics.
Due to the complete loss of the quantum coherence in the qubit eigenbasis, the von Neumann entropy of the classicalized state, whose density matrix is diagonal in the qubit eigenbasis, is the Shannon entropy and measures the information lost by classicalization \cite{EPL1}.

In our formulation of the classicalized HTN (cHTN) in Ref. \cite{JHAP1}, we fixed the geometry of the bulk spacetime, in which a massive particle is present, to the cHTN of the ground state as the background discrete hyperbolic geometry \cite{RINP} and treated this particle in its non-relativistic, Euclidean regime.
Then, we obtained the bulk quantum mechanical path integral of the particle in the Lorentzian regime from the bulk classical stochastic process of the particle in the Euclidean regime by the inverse Wick rotation \cite{JHAP1,JHAP4}.

In Ref. \cite{EPL2}, the action of the cHTN in the classicalized ground state $|\psi\ra=(|\psi\ra,{\cal A})$ is given by
\begin{equation}
I[|\psi\ra]=-\hbar H[|\psi\ra]\;,\label{eq:I1}
\end{equation}
where $H$ denotes the Shannon entropy in nats.
This action is based on the holographic principle, which equates the boundary information and the bulk degrees of freedom \cite{Hol1,Hol2,Hol3}, and consists of the bulk degrees of freedom (i.e., the boundary information) {\it lost} by the classicalization of the ground state of the boundary CFT$_2$.
Each bulk degree of freedom has the spin-half action $\hbar$ \cite{EPL2}.

To incorporate time dependence in the formulation of the cHTN in the Lorentzian regime, in Ref. \cite{JHAP3}, we classicalized real time, too, and considered the special relativistic real-time mixed state (i.e., time ensemble) of a generic quantum pure state $|\psi(t)\ra$ of the boundary CFT$_2$ in the representation of the Lorentzian boundary conformal symmetry,
\begin{equation}
|\psi\rra_L{}_L\lla \psi|=\int_{{\cal T}} |\psi(t)\ra \la \psi(t)|d\mu (t)\;,\label{eq:te}
\end{equation}
for an absolutely continuous temporal measure $d\mu(t)$ of the density matrix $|\psi\rra_L{}_L\lla \psi|$ and the boundary total temporal domain ${\cal T}$ where $|\psi(t)\ra$ is defined \cite{JHAP3,AnnMath}.\footnote{To take the square root of the density matrix (\ref{eq:te}), we need an auxiliary temporal sector \cite{JHAP3}.
In particular, if $|\psi\rra_L$ is the ground state, $|\psi\rra_L{}_L\lla \psi|=|\psi\ra\la \psi|\otimes \widehat{1}$ holds for $\widehat{1}=|{\cal T}\ra\la {\cal T}|$ \cite{JHAP3}.}
This measure $d\mu(t)$ of the time ensemble satisfies the normalization condition on the trace of the density matrix
\begin{equation}
\int_{{\cal T}} d\mu(t)=1\;.
\end{equation}
Here, a definite form of the time ensemble appears in quantum statistical mechanics as the quantum mixed state
\begin{equation}
\lim_{T\to \infty}\frac{1}{2T}\int_{t=-T}^T|\psi(t)\ra \la \psi(t)|dt\label{eq:vN}
\end{equation}
of the state vectors $|\psi(t)\ra$ under the unitary time evolution of $|\psi(t)\ra$, in the Hilbert space, governed by a Hamiltonian.
This time ensemble was introduced by von Neumann in Ref. \cite{vN} to define the long-time average of observables $\widehat{{\cal O}}$ by
\begin{equation}
\overline{\la \widehat{{\cal O}}\ra}\equiv \lim_{T\to \infty}\frac{1}{2T}\int_{t=-T}^T\la \psi(t)|\widehat{{\cal O}}|\psi(t)\ra dt
\end{equation}
in his proof of the quantum ergodicity of an isolated quantum system described by the state vector $|\psi(t)\ra$.
In this time ensemble, $dt/2T$ constitutes the temporal realization probability of a state vector $|\psi(t)\ra$ in the Hilbert space.

In Ref. \cite{JHAP3}, the Lorentzian action of the cHTN is given by
\begin{eqnarray}
I_L[|\psi\rra_L]=-\hbar H[|\psi\rra_L]\label{eq:IE0}
\end{eqnarray}
for the classicalized mixed state $|\psi\rra_L=(|\psi\rra_L,{\cal A})$.
In terms of the time ensemble, it incorporates the time dependence of $|\psi\rra_L$ in the action (\ref{eq:I1}).
To illustrate the meaning of the Shannon entropy in this action, let us consider the discrete time ensemble
\begin{equation}
|\psi\rra_L{}_L\lla \psi|=\sum_k p_k|\psi_k\ra\la \psi_k|\label{eq:pk}
\end{equation}
for the temporal realization probabilities $p_k$ of cHTN mixed states $|\psi_k\ra=(|\psi_k\ra,{\cal A})$, which have support on orthogonal subspaces of ${\mathfrak H}$.
Here, each probability $p_k$ is defined by
\begin{equation}
p_k\equiv \int_{{\cal T}_k}d\mu(t)
\end{equation}
for the boundary total temporal domain ${\cal T}_k$ occupied by the cHTN mixed state $|\psi_k\ra$ such that the normalization condition on the trace of the density matrix (\ref{eq:pk}), that is,
\begin{equation}
{\cal T}=\coprod_k {\cal T}_k\;,\label{eq:Tk}
\end{equation}
holds for the disjoint union with respect to ${\cal T}_k$.
Now, the Shannon entropy of $|\psi\rra_L$ is \cite{NC}
\begin{equation}
H[|\psi\rra_L]=\sum_k p_k (-\ln p_k +S_{\rm v.N.}[|\psi_k\ra])\label{eq:NC}
\end{equation}
for the von Neumann entropy in nats ($S_{\rm v.N.}$).
In Eq. (\ref{eq:NC}), the first and second terms are the entropy of the time ensemble and the temporal average of the entropy of the cHTN, respectively.

Here, we make a remark on the time ensemble.
\begin{enumerate}
\item[(A)] In the Euclidean regime, we regard $-\hbar \ln p_k$ (here, $-\ln p_k$ is information) as the informational Euclidean action of the boundary temporal domain ${\cal T}_k$: ergodicity of the cHTN is adopted as a principle.
Then, from the entropic Euclidean action of the cHTN \cite{JHAP3}, we obtain the bulk imaginary-time path integral of the discrete geometry of $|\psi_k\ra$ in the {\it off-shell} treatment of $|\psi_k\ra$.
\end{enumerate}

When the boundary total temporal domain ${\cal T}$ is an infinitesimal time interval that contains $t=0$ (corresponding to the label $k=0$), $p_0=1$ holds in Eq. (\ref{eq:NC}): the time ensemble $|\psi\rra_L$ is a pure time-ensemble.
In this case of $|\psi\rra_L$, the maximum entropy with respect to $|\psi_0\ra$ is
\begin{equation}
H_{\rm max}[|\psi_0\ra]=\sigma A_{{\rm TN},0}\label{eq:max}
\end{equation}
for the von Neumann entropy of the cHTN per {\it site} (i.e., per disentangler) $\sigma$ in nats and the number of sites $A_{{\rm TN},0}$ defined for the cHTN mixed state $|\psi_0\ra$ \cite{EPL2}.
Here, $|\psi_0\ra$ has discrete hyperbolic geometry \cite{RINP}, and $\sigma$ is the von Neumann entropy of a classicalized Bell state in nats and is $\ln 2$ \cite{EPL2} in the exact strong-coupling limit of the boundary CFT$_2$.

From here, we assume a relativistic particle with a non-zero mass $M$ in the bulk spacetime.
In this setup, the Lorentzian action is given by
\begin{eqnarray}
I_L[|\psi\rra_L,\gamma_t]=-\hbar H[|\psi\rra_L]+S_L[\gamma_t]\;,\label{eq:IE1}
\end{eqnarray}
where $S_L$ is the relativistic action of the real-time world line $\gamma_t$ of the particle $M$ and is given by \cite{Landau}
\begin{equation}
S_L[\gamma_t]=-Mc^2\int_{\gamma_t} d\tau_{\gamma_t}\label{eq:SL0}
\end{equation}
for the bulk real proper time $\tau$.
In the discrete hyperbolic geometry of the cHTN, the bulk real proper time $\tau$ is obtained from the $SO(2,2)$ isometry of the Lorentzian bulk spacetime in the absence of the length scale variable of the cHTN \cite{JHAP3}.
Note that $-S_L/\hbar$ represents entropy because $H$ on equal footing is entropy: see Eq. (\ref{eq:SL}).

We decompose the boundary total time interval to be considered into the disjoint union of infinitesimal time intervals.
Then, the {\it on-shell} equation of the cHTN in the presence of a non-zero mass $M$ is obtained from the variation of the action with respect to the bulk real proper time $\tau$ at the site of the mass:
\begin{equation}
-\hbar\frac{1}{d^2A}d_\tau d^2H_{\rm max}+d_\tau S_L=0\;,\label{eq:Theta0}
\end{equation}
where the action is temporally globally minimized at $-\hbar H_{\rm max}$ (i.e., the entropy is temporally globally maximized at $H_{\rm max}$).
In Eq. (\ref{eq:Theta0}), the cHTN part is {\it per} site and its action is minimized at $-\hbar H_{\rm max}$ in each infinitesimal time interval.
$d^2A$ is the infinitesimal dimensionless area of the cHTN around the site of the mass.
From Eqs. (\ref{eq:max}) and (\ref{eq:SL0}), we obtain
\begin{eqnarray}
d_\tau d^2H_{\rm max}&=&\sigma \theta d^2A d\tau\;,\label{eq:sub1}\\
d_\tau S_L&=&-Mc^2 d\tau\;,\label{eq:sub2}
\end{eqnarray}
respectively.
Here,
\begin{eqnarray}
\theta=\frac{1}{d^2A}\frac{d}{d\tau} d^2A
\end{eqnarray}
is the real-proper-time expansion of $d^2A$.

Substituting Eqs. (\ref{eq:sub1}) and (\ref{eq:sub2}) into Eq. (\ref{eq:Theta0}), we obtain our main result:
\begin{equation}
-\sigma \hbar \theta=Mc^2\;.\label{eq:Theta}
\end{equation}
Note that Eq. (\ref{eq:Theta}) is the relativistic on-shell equation of the cHTN in the Lorentzian regime, and the discrete hyperbolic geometry of the cHTN incorporates the negative fundamental cosmological constant of AdS$_3$ spacetime in general relativity.

In Eq. (\ref{eq:Theta}), $\sigma$ and $Mc^2$ represent entropy production and energy present at the site of the mass $M$, respectively.
So, the energy quantum $-\hbar \theta$ can be interpreted as the thermal energy of the cHTN
\begin{equation}
-\hbar \theta =k_BT\label{eq:thermo}
\end{equation}
at a finite temperature
\begin{equation}
T=\frac{Mc^2}{\sigma k_B}
\end{equation}
in the presence of a non-zero mass $M$ in the bulk spacetime.

Here, we make an observation: the combination
\begin{equation}
\theta d\tau =\frac{1}{d^2A}d_\tau d^2A\label{eq:theta}
\end{equation}
is scale-invariant.
Thus, by rescaling the discrete length, this combination can be defined at the {\it top tensor} \cite{Vidal2} of the cHTN.
Then, the inverse renormalization group (RG) direction is defined from the top tensor of the cHTN.
For this direction, the combination (\ref{eq:theta}) means temporal relabeling of the inverse RG steps such as
\begin{equation}
\theta d\tau=(\ln 2) d_\tau n\label{eq:gauge}
\end{equation}
for the inverse RG step, $n$, counted from the top tensor.
Note that Eq. (\ref{eq:gauge}) is always negative.

In the interpretation (\ref{eq:thermo}), we equally divide the entropy
\begin{equation}
-\frac{d_\tau S_L}{\hbar}=\sigma \frac{k_BT d\tau}{\hbar}\label{eq:SL}
\end{equation}
at the top tensor into entropies $\sigma k_BT^U_n d\tau_n/\hbar$ for the Unruh temperature $T^U_n$ and the bulk real proper time $\tau_n$ at the inverse RG step $n$ as
\begin{equation}
k_BT^U_n d\tau_n =\frac{k_BTd \tau}{N_n}\label{eq:Unruh1}
\end{equation}
for the number of sites $N_n$ at the inverse RG step $n$.
For $T^U_n$, we define the {\it Lorentzian on-shell proper acceleration}, of magnitude $a_{L,n}$, of another non-zero mass at the inverse RG step $n$ by the Unruh effect \cite{Review0,Davies,Unruh1,Sewell,Unruh2,Harlow2}:
\begin{equation}
T^U_n=\frac{\hbar a_{L,n}}{2\pi ck_B}\;.\label{eq:Unruh2}
\end{equation}
Equating the relations (\ref{eq:thermo}), (\ref{eq:gauge}), (\ref{eq:Unruh1}), and (\ref{eq:Unruh2}), $a_{L,n}$ is proportional to the proper-time derivative $dn/d\tau_n$ and the inverse length factor $1/N_n$ at step $n$.

The following three properties of $a_L$ follow from Eq. (\ref{eq:Theta}):
\begin{enumerate}
\item[(i)] $\theta d\tau$ is scale-invariant.
The unity spatial codimension of the inverse RG steps indicates the ``{\it inverse-square law}'' of $a_L$ in three spacetime dimensions, including the bulk redshift effect.

\item[(ii)] $\theta d\tau$ is proportional to the mass $M$.
From this fact, $a_L$ is also proportional to $M$.

\item[(iii)] The left-hand side of Eq. (\ref{eq:Theta}) has a negative sign.
Since the inverse RG step $n$ always has a negative proper-time derivative, the universal attractivity of $a_L$ follows.

\end{enumerate}

Next, we add four remarks.
\begin{enumerate}
\item[(B)] In Eq. (\ref{eq:Theta}), the action of the mass $M$ on the cHTN is scale-invariant: $d^2A$ temporally contracts in a self-similar way with the characteristic proper time $\sigma \hbar/ Mc^2$.

\item[(C)]
$a_{L,n}$ is the inverse proper-time length scale per site \cite{Review0,Harlow2} at the inverse RG step $n$.
As derived above, $a_{L,n}$ is proportional to both $dn/d\tau_n$ and $1/N_n$.

\item[(D)]
We set the cut-off proper length of the cHTN as $\Lambda$.
Then, the number of redefined sites $N_n$ at a fixed step $n$ counted from the top tensor by $\Lambda$ is independent of the choice of $\Lambda$.

\item[(E)]
The MERA describes a real-time evolution and has a causal light-cone structure \cite{Vidal1,Vidal2}.
Thus, the MERA cannot be regarded as a Euclidean time slice of the AdS$_3$ spacetime \cite{Review0}.
However, after classicalization, the MERA recovers the discrete hyperbolic spatial metric \cite{RINP}.
\end{enumerate}

Finally, we consider the flat-spacetime limit of the cHTN \cite{Flat,JHAP2}.
In this limit, the inverse RG step from a non-zero mass $M$ to null infinity is infinitely divided in a finite (contracted) AdS scale \cite{JHAP2}, and these divisions are spatially concentric circles due to the $SO(2)$ isometry.
Then, the Lorentzian on-shell acceleration $a_L$ of another non-zero mass separated from the mass $M$ is determined in the form of the three-dimensional gravitational acceleration, except for the indefiniteness of the cut-off proper length, by the above three properties (i), (ii), and (iii) of $a_L$.


\begin{thebibliography}{99}
\bibitem{Hol1}G. 't Hooft, arXiv:gr-qc/9310026.
\bibitem{Hol2}L. Susskind,
J. Math. Phys. {\bf 36}, 6377 (1995).
\bibitem{Hol3}R. Bousso,
Rev. Mod. Phys. {\bf 74}, 825 (2002).
\bibitem{AdSCFT1}J. M. Maldacena, 
Adv. Theor. Math. Phys. {\bf 2}, 231 (1998).
\bibitem{AdSCFT2}S. S. Gubser, I. R. Klebanov and A. M. Polyakov,
Phys. Lett. B {\bf 428}, 105 (1998).
\bibitem{AdSCFT3}E. Witten,
Adv. Theor. Math. Phys. {\bf 2}, 253 (1998).
\bibitem{AdSCFT4}O. Aharony, S. S. Gubser, J. M. Maldacena, H. Ooguri and Y. Oz,
 Phys. Rep. {\bf 323}, 183 (2000).
\bibitem{Nastase}H. N{\u{a}}stase,
{\it Introduction to the AdS/CFT Correspondence}
(Cambridge University Press, Cambridge, 2015).
\bibitem{Swingle1}B. Swingle, 
Phys. Rev. D {\bf 86}, 065007 (2012).
\bibitem{Vidal1}G. Vidal, 
Phys. Rev. Lett. {\bf 99}, 220405 (2007).
\bibitem{Vidal2}G. Vidal, 
Phys. Rev. Lett. {\bf 101}, 110501 (2008).
\bibitem{RT1}S. Ryu and T. Takayanagi, 
Phys. Rev. Lett. {\bf 96}, 181602 (2006).
\bibitem{RT2}S. Ryu and T. Takayanagi, 
J. High Energy Phys. {\bf 08}, 045 (2006).
\bibitem{HRT}V. E. Hubeny, M. Rangamani and T. Takayanagi, 
J. High Energy Phys. {\bf 07}, 062 (2007).
\bibitem{RT3}M. Rangamani and T. Takayanagi, 
Lect. Notes Phys., Vol. {\bf 931} Springer (2017).
\bibitem{Review0}B. Chen, B. Czech and Z. Wang,
Rep. Prog. Phys. {\bf 85}, 046001 (2022).
\bibitem{Matsueda}H. Matsueda, M. Ishibashi and Y. Hashizume,
Phys. Rev. D {\bf 87}, 066002 (2013).
\bibitem{Review1}N. Bao, C. Cao, S. M. Carroll, A. Chatwin-Davies and N. Hunter-Jones,
Phys. Rev. D {\bf 91}, 125036 (2015).
\bibitem{Swingle2}B. Swingle,
Annu. Rev. Condens. Matter Phys. {\bf 9}, 345 (2018).
\bibitem{Review3}A. Jahn and J. Eisert,
Quantum Sci. Technol. {\bf 6}, 033002 (2021).
\bibitem{EPL1}E. Konishi, 
EPL {\bf 129}, 11006 (2020).
\bibitem{EPL2}E. Konishi,
 EPL {\bf 132}, 59901 (2020), arXiv:1903.11244 [quant-ph].
 \bibitem{JHAP1}E. Konishi,
JHAP {\bf 1}, (1) 47-56 (2021).
\bibitem{ZurekPT}W. H. Zurek,
Phys. Today {\bf 44}, (10) 36-44 (1991).
\bibitem{RINP}E. Konishi,
 Results in Physics {\bf 19}, 103410 (2020).
\bibitem{JHAP4}E. Konishi,
JHAP {\bf 3}, (1) 31-38 (2023).
\bibitem{JHAP3}E. Konishi,
JHAP {\bf 2}, (4) 1-10 (2022).
\bibitem{AnnMath}J. von Neumann,
Ann. Math. {\bf 50}, 2 (1949).
\bibitem{vN}J. von Neumann,
Eur. Phys. J. H {\bf 35}, 201 (2010).
\bibitem{NC}M. A. Nielsen and I. L. Chuang,
{\it Quantum Computation and Quantum Information} (Cambridge University Press, Cambridge, 2000).
\bibitem{Landau}L. D. Landau and E. M. Lifshitz,
{\it The Classical Theory of Fields}
(Butterworth-Heinemann, Oxford, 1975).
\bibitem{Davies}P. C. W. Davies,
J. Phys. A {\bf 8}, 609 (1975).
\bibitem{Unruh1}W. G. Unruh, 
Phys. Rev. D {\bf 14}, 870 (1976).
\bibitem{Sewell}G. L. Sewell,
Ann. Phys. {\bf 141}, 201 (1982).
\bibitem{Unruh2}L. C. B. Crispino, A. Higuchi and G. E. A. Matsas,
Rev. Mod. Phys. {\bf 80}, 787 (2008).
\bibitem{Harlow2}D. Harlow,
Rev. Mod. Phys. {\bf 88}, 015002 (2016).
\bibitem{Flat}A. Bagchi and R. Fareghbal,
J. High Energy Phys. {\bf 10}, 092 (2012).
\bibitem{JHAP2}E. Konishi,
JHAP {\bf 2}, (3) 71-80 (2022).
\end{thebibliography}
\end{document}